\documentclass[12pt]{amsart}
\usepackage{amssymb,latexsym,amsfonts,amsmath,amsthm,verbatim,eucal,hyperref}
\def\PP{\mathbb{P}}
\def\EE{\mathbb{E}}
\def\RR{\mathbb{R}}
\def\NN{\mathbb{N}}
\def\ZZ{\mathbb{Z}}

\def\1{\mathbf{1}}
\def\eps{\varepsilon}

\def\tr{\operatorname{tr}}

\def\To{\longrightarrow}

\theoremstyle{plain}
\newtheorem{thm}{Theorem}[section]
\newtheorem{lemma}[thm]{Lemma}
\newtheorem*{lemma*}{Lemma*}
\newtheorem{prop}[thm]{Proposition}
\newtheorem*{prop*}{Proposition}
\theoremstyle{remark}
\newtheorem{rmk}[thm]{Remark}

\theoremstyle{definition}
\newtheorem{cor}[thm]{Corollary}
\newtheorem{ex}[thm]{Example}
\newtheorem{dfn}[thm]{Definition}

\begin{document}

\title[Random matrices, non-backtracking walks \dots]
      {Random matrices, non-backtracking walks, and orthogonal polynomials}
\author[Sasha Sodin]
       {Sasha Sodin}
\address{School of Mathematics,
         Raymond and Beverly Sackler Faculty of Exact Sciences,
         Tel Aviv University, Tel Aviv, 69978, Israel.}
\email{sodinale@tau.ac.il}
\date{\today}

\begin{abstract}
Several well-known results from the random matrix theory, such as
Wigner's law and the Marchenko--Pastur law, can be interpreted (and
proved) in terms of non-backtracking walks on a certain graph.
Orthogonal polynomials with respect to the limiting spectral measure
play a r\^ole in this approach.
\end{abstract}

\maketitle{}

\section{Introduction}

Our goal is to explain a unified approach to the proofs of several
well-known theorems in the spectral theory of random matrices
and random graphs. Some of these results are formulated further
in the introduction; striving to make the main idea as clear as
possible, we restrict ourselves to paradigmatic examples.
In particular, we only consider Bernoulli random matrices, although
most proofs can be adapted to arbitrary random variables under mild
assumptions on tail decay.

The method may be seen as a modification of the {\em moment method};
in the latter, used extensively since Wigner, spectral properties of
a matrix $M$ are extracted from the traces $\tr M^k$ of powers
of $M$. Instead, we propose to estimate $\tr P_k(M)$,
where $P_k$ are orthogonal polynomials with respect to a certain measure
$\sigma$, which is the candidate for limiting spectral measure.
Perhaps surprisingly, these numbers have, in some cases, a simple
combinatorial interpretation, in terms of non-backtracking walks
(see Subsection~\ref{s:fin}) on an appropriate graph.

One can also start from a linear recurrent relation of order two for the
number of non-backtracking walks. Then a measure $\sigma$ appears
from the correspondence between Jacobi (tridiagonal) matrices and measures
on $\RR$. This classical correspondence involves the orthogonal polynomials
$P_k$ with respect to $\sigma$, that satisfy the same recurrent relation.
In fact, we will see (see e.g.\ Lemma~\ref{mainlemma}) that the matrix
$P_k(M)$ is closely related to non-backtracking walks of length $k$.

Now it is natural to guess that $\sigma$ is the limiting spectral measure.
We show that this is the case if the traces $\tr P_k(M)$ do not grow too fast;
the proof is based on an analytic lemma (cf.\ Subsection~\ref{s:cms}).
The combinatorial estimates (in Section~\ref{S:cnt}) allow to bound these
traces, for the examples that we consider.

Although orthogonal polynomials do not appear explicitly in the work of
Bai and Yin on the smallest singular value of a random covariance
matrix \cite{BY}, the present note (as well as part of the previous
work \cite{AFMS}) started from an attempt to understand and generalise
their proof.

Similar ideas emerged also in the spectral graph theory, starting from the
work of McKay \cite{MK1,MK2}. McKay derived an expression for the number
of non-backtracking walks on a graph in terms of certain polynomials of
the adjacency matrix from a certain recurrent relation and applied it to
study the spectral measure of $d$-regular graphs; Friedman \cite{F} applied
it to study the spectral gap of random graphs. Li and
Sol\'e \cite{LS} noted that these are exactly the orthogonal polynomials
with respect to the Kesten--McKay measure (\ref{km_meas}), and suggested
to consider more general measures of the Bernstein--Szeg\H{o} class (see
Section~\ref{s:bs}). They also used the Chebyshev--Markov--Stieltjes
inequalities (cf.\ Subsection \ref{s:mk}). Related methods were developed
by Brooks \cite{B} and Serre \cite{S}.

We try to emphasise the applications to matrices other than the adjacency
matrix of a graph, and especially -- to random matrices.

{\it Acknowledgement.} I am grateful to my supervisor Vitali Milman for his support
and useful discussions, and for urging me to write this note. The mini-courses on
Random Matrix Theory taught by Leonid Pastur and Mariya Shcherbina (in Vienna and
Paris) greatly improved my understanding of this field. My father helped me find
the way in the literature on the problem of moments. Bo$'$az Klartag, Michel Ledoux, 
Brendan McKay, and Paul Nevai have kindly commented on a preliminary version of this note.
I thank them all very much.

\subsection{Two definitions and notation}

\begin{dfn}
Let $M$ be an $n \times n$ symmetric matrix; let
\[ \lambda_1(M) \leq \lambda_2(M) \leq \cdots \leq \lambda_n(M)\]
be the eigenvalues of $A$.
The measure $\mu_M$,
\begin{equation}
\mu_M(S) = \# \left\{ 1 \leq j \leq n \, \mid \,
    \lambda_j(M) \in S \right\}, \quad S \subset \RR~,
\end{equation}
is called the {\em spectral measure} of $M$.
\end{dfn}

\begin{dfn}
Let $\mu, \nu$ be two probability measures on $\RR$. The Kolmogorov
distance between $\mu$ and $\nu$ is defined as
\[ d_\text{K}(\mu, \nu)
    = \sup_{x \in \RR} \left| \mu(-\infty, x] - \nu(- \infty, x] \right|~.\]
\end{dfn}

{\em Notation:} Unless otherwise specified, $C, C_1, C_2, C', c, c', \cdots$
denote positive constants not depending on any parameters of the problem.
Usually, upper case $C$ stand for a big constant, and lower case $c$ - for
a small constant.

\subsection{Symmetric random matrices}\label{f:w}

For $n \in \NN$, let $A$ be a symmetric $n \times n$ random
matrix, such that
\begin{equation}\label{d:w}
\begin{cases}
\text{$A_{uv}$ are independent for $1 \leq u \leq v \leq n$}, \\
\PP \left\{ A_{uv} = -1/(2\sqrt{n}) \right\}
    = \PP \left\{ A_{uv} = 1/(2\sqrt{n}) \right\} = 1/2.
\end{cases}
\end{equation}

\begin{thm}[Wigner's law]
As $n \to \infty$, the random measures $\mu_A$ converge (weakly,
in distribution) to a deterministic measure $\sigma_\text{W}$
supported on $[-1, 1]$;
\[ d\sigma_\text{W}(x) = \frac{2}{\pi} \sqrt{1 - x^2} \, dx~.\]
\end{thm}
The measure $\sigma_\text{W}$ is called the Wigner measure.

\begin{rmk}[Precise meaning of convergence]\label{top}
The space $\mathcal{M}(\RR)$ of measures on $\RR$
is equipped with the weak topology. For every $n \in \NN$,
the measure $\mu_A$ is a random element of $\mathcal{M}(\RR)$;
its distribution is a probability measure on $\mathcal{M}(\RR)$.
In Wigner's law, these distributions converge (weakly) to the
distribution $\delta_{\sigma_\text{W}}$ supported on a single point
$\sigma_\text{W} \in \mathcal{M}(\RR)$.
\end{rmk}

\begin{thm}[F\"uredi--Koml\'os \cite{FK}]\label{t:fk}
As $n \to \infty$, the operator norm
\[ \| A \| = \max(|\lambda_1|, |\lambda_n|) \]
of $A$ converges (in distribution) to $1$.
\end{thm}

Wigner's theorem (above) implies that
\[ \PP \left\{ \| A \| \leq 1 - \eps \right\} \To 0 \]
for any $\eps > 0$. As for the complimentary inequality,
we prove a stronger fact:

\begin{thm}[A.~Boutet de Monvel and M.~Shcherbina \cite{BS}]\label{t:bs}
For some (universal) constants $c, \alpha_1, \alpha_2, \alpha_3 > 0$,
\begin{equation}
\PP \{ \| A \| \geq 1 + \eps \}
    \leq \exp(-c n^{\alpha_1} \eps^{\alpha_2})~,
\end{equation}
provided that
\[ n^{-\alpha_3} \leq \eps \leq 1~.\]
\end{thm}

\subsection{Random covariance matrices}\label{f:mp}

For $n \leq N$, let $B$ be an $n \times N$ random matrix (that
is, $B: \RR^N \to \RR^n$), so that
\begin{equation}\label{d:mp}\begin{cases}
\text{$B_{uv}$ are independent for $1 \leq u \leq n$, $1 \leq v \leq N$}, \\
\PP \left\{ B_{uv} = -1/\sqrt{N} \right\}
    = \PP \left\{ B_{uv} = 1/\sqrt{N} \right\} = 1/2.
\end{cases}\end{equation}

Now we are interested in the eigenvalues
\[ 0 \leq \lambda_1 \leq \cdots \leq \lambda_n \]
of the (symmetric) matrix $C = BB^t$.

\begin{thm}[Marchenko--Pastur \cite{MP}]
If $n,N \To \infty$ so that
\[ n/N \To \xi \in (0, 1]~,\]
the spectral measure $\mu_C$ converges (weakly, in distribution) to a
deterministic measure
$\sigma^\xi_\text{MP}$ supported on $[(1-\sqrt{\xi})^2, (1+\sqrt{\xi})^2]$;
\[ d\sigma_\text{MP}^\xi(x)
    = \frac{1}{2\pi\xi x}
        \sqrt{\left(x - (1-\sqrt{\xi})^2\right)
        \left((1+\sqrt{\xi})^2 - x\right)} \, dx~.\]
\end{thm}

The measure $\sigma^\xi_\text{MP}$ is called the Marchenko--Pastur measure.

\begin{thm}[Geman \cite{G}, Bai--Yin \cite{BY}]
If $n,N \To \infty$ so that
\[ n/N \To \xi \in (0, 1]~,\]
the smallest eigenvalue of $C$ converges (in distribution) to
$(1-\sqrt{\xi})^2$, and the largest - to $(1+\sqrt{\xi})^2$.
\end{thm}

\begin{rmk}
The convergence of the largest eigenvalue was proved by Geman, and of
the smallest - by Bai and Yin.
\end{rmk}

Similarly to the previous subsection,
\[ \PP \left\{ \lambda_1(C) \geq (1 - \sqrt{\xi})^2 + \eps \right\} \To 0 \]
and
\[ \PP \left\{ \lambda_n(C) \leq (1 + \sqrt{\xi})^2 - \eps \right\} \To 0 \]
by the Marchenko--Pastur theorem. As for the complementary inequalities, we prove
the following:

\begin{thm}[\cite{AFMS}]\label{afms}
For some (universal) constants $c, \beta_1, \beta_2, \beta_3 > 0$,
\begin{eqnarray}
\PP \left\{ \lambda_1(C) \leq (1-\sqrt{\xi})^2 - \eps \right\}
    &\leq& \exp(- c n^{\beta_1} \eps^{\beta_2})~, \label{eq:afms.1}\\
\PP \left\{ \lambda_n(C) \geq (1+\sqrt{\xi})^2 + \eps \right\}
    &\leq& \exp(- c n^{\beta_1} \eps^{\beta_2})~, \label{eq:afms.2}
\end{eqnarray}
provided that
\[n^{-\beta_3} \leq \eps \leq 1~.\]
\end{thm}

\subsection{Adjacency matrix of a random $d$-regular graph}\label{f:mk}

Fix $d \geq 3$; let $G = (V, E)$ be a random $d$-regular graph on $n$
vertices. That is, $G$ is picked uniformly from the collection of all
graphs $G = (V, E)$ such that $\# V = n$ and
\[ \# \left\{ u \in V \, \mid \, (u, v) \in E \right\} = d
    \quad \text{for every $v \in V$}~.\]

Let $A(G)$ be the adjacency matrix of $G$; that is,
\[ {A(G)}_{uv} = \begin{cases}
1, &(u, v) \in E \\
0, &\text{otherwise}~.
\end{cases}\]

\begin{thm}[McKay]
The spectral measure $\mu_{A(G)}$ converges (weakly, in distribution,
as $n \to \infty$) to a deterministic measure
$\sigma_\text{KM}$ supported on
\[ [-2 \sqrt{d-1}, 2 \sqrt{d-1}]~; \]
\begin{equation}\label{km_meas}
d\sigma_\text{KM}(x)
    = \frac{d}{2\pi}
        \frac{\sqrt{4(d-1)-x^2}}{d^2 - x^2} \, dx~.
\end{equation}
\end{thm}

The measure $\sigma_\text{KM}$ is called the Kesten--McKay measure.

\subsection{A guide to the next sections}

In Subsection~\ref{s:not} we introduce the general framework that unites
all the problems listed above. In Subsection~\ref{s:ex} we focus on an
example,-- the infinite $d$-regular tree,-- that should clarify the meaning
of the Kesten--McKay measure, and also hint the main idea in the proofs of
all the theorems. Lemma~\ref{mainlemma} in Subsection~\ref{s:fin} relates
the spectral properties of the matrices in study to certain combinatorial
quantities.

We apply it in Subsection~\ref{s:mk} to prove McKay's theorem, and in
Subsection~\ref{s:w} -- to prove Wigner's theorem. In Subsection~\ref{s:mp}
we sketch the proof of the Marchenko--Pastur theorem. The bounds on extremal
eigenvalues are the subject of Section~\ref{S:norm}.

Section~\ref{S:orth} recalls some properties of orthogonal polynomials
with respect to measures that appear in this note. In Section~\ref{S:cnt}
we prove the combinatorial estimates used in the proofs of the theorems
on random matrices. These two sections contain the technical results
that we use elsewhere.

\section{Spectral measure: limit theorems}\label{S:sm}

\subsection{Matrices on graphs}\label{s:not}

Let $G = (V, E)$ be a graph (with vertices $V$ and edges $E$). A (symmetric)
$V \times V$ matrix $M$ is called a {\em (symmetric) sign matrix} on $G$ if
\[ M_{uv} = \begin{cases}
\pm 1, & (u,v) \in E \\
0, &(u, v) \notin E \end{cases}. \]

\begin{ex} If
\[ M_{uv} = \begin{cases}
+ 1, & (u,v) \in E \\
0, &(u, v) \notin E \end{cases}, \]
$M$ is the adjacency matrix $A(G)$ of $G$.
\end{ex}

If the degree of every vertex is finite,-- that is,
\[ \deg(v) = \# \left\{ v \in V \, \mid \, (u, v) \in E \right\}
    < + \infty\]
for every $v \in V$,-- the matrix $M$ defines a symmetric operator on a dense
subspace of $L_2(V)$. If moreover the degrees are uniformly bounded by a
number $D$, $M$ is self-adjoint and $\|M\| \leq D$.

We are mainly interested in finite graphs ($\# V < + \infty$); however, it
will be convenient to have the definitions in this generality.

Let us recall the spectral theorem for self-adjoint operators (see Akhiezer
and Glazman \cite{AG}).

\begin{dfn} A family of projectors $\{ E_t \, \mid \, -D \leq t \leq +D \}$
is called a {\em resolution of identity} if
\begin{enumerate}
\item $E_{-D}= 0$, $E_{+D} = \1$
\item $E_t E_t' = E_{\min(t, t')}$
\item $\displaystyle{\lim_{t \to t'-0}} E_t = E_{t'}$.
\end{enumerate}
\end{dfn}

For our operator $M$, there exists a resolution of identity such that all
$E_t$ commute with $M$ and
\begin{equation}\label{st1}
p(M) = \int_{-D}^D p(t) dE_t
\end{equation}
for any polynomial $p$.

The (operator-valued) measure $dE_t$ is called the {\em spectral measure} of $M$.
In some important cases the (real) measure $d\langle E_t \delta_v, \delta_v \rangle$
does not depend on the choice of a vertex $v \in V$ (here $\delta_v(u) = \delta_{uv}$
for $u,v \in V$). In this case, we also call it the spectral measure of $M$ (more
general definitions are available for $M = A(G)$; see Grigorchuk and \.Zuk \cite{GZ}
and references therein).

\subsection{Main example}\label{s:ex}

Denote by $H_d = (V_d, E_d)$ the (infinite) $d$-regular tree ($d \geq 3$); let
$M$ be a symmetric sign matrix on $H_d$. According to (\ref{st1}),
\[ \langle p(M) f, f \rangle = \int_{-d}^d p(t) d \langle E_t f, f \rangle \]
for any polynomial $p$ and any $f \in L_2(V_d)$, and in particular
\begin{equation}\label{st2}
\langle p(M) \delta_u, \delta_u \rangle
    = \int_{-d}^d p(t) d \langle E_t \delta_u, \delta_u \rangle~.
\end{equation}
Note that the measures
$d\langle E_t \delta_u, \delta_u \rangle$ do not depend on $u$ (because
of homogeneity). In fact, these measures also do not depend on $M$. The
following fact is essentially due to Kesten \cite{K}:

\begin{prop}\label{me} The measures
$d\langle E_t \delta_u, \delta_u \rangle$ are equal to the Kesten--McKay
measure $\sigma^d_\text{KM}$.
\end{prop}

\begin{proof}
Define a sequence of polynomials
\[ (p_k)_{k \in \ZZ_+} = (p_{k,d})_{k \in \ZZ_+}, \quad \deg p_k = k~:\]
\begin{equation}\label{pk}
\begin{cases}
&p_0(t) = 1, \quad p_1(t) = t\Big/\sqrt{d}, \\
&p_2(t) = t^2\Big/\sqrt{d(d-1)} - \sqrt\frac{d}{d-1}, \\
&p_{k+1}(t) = t \, p_k(t)\Big/\sqrt{d-1} - p_{k-1}(t) \quad (k = 2, 3, \cdots)~.
\end{cases}
\end{equation}

\begin{lemma}\label{l1}
\[ \langle p_k(M) \delta_u, \, \delta_u \rangle = 0
    \quad \text{for $k = 1,2,3\cdots$.} \]
\end{lemma}
As we shall see (in Lemma~\ref{mainlemma}, from which our lemma follows), 
this equality expresses the fact that ``there are no cycles in $H_d$''. 
Now we need one more property of the polynomials $p_k$; for proof, see 
Remark~\ref{rmkKM} in Section~\ref{S:orth} (and the discussion preceding it).

\begin{lemma}\label{l2} The polynomials $p_k$ are orthogonal
with respect to the measure $\sigma_\text{KM}$:
\[ \int_{-d}^d p_k(t) p_l(t) \, d\sigma^d_\text{KM}(t) = \delta_{kl}~,
\quad k,l \in \ZZ_+~.\]
\end{lemma}

In view of  (\ref{st2}) and Lemma~\ref{l1},
\[ \int_{-d}^d p_k(t) d \langle E_t \delta_u, \delta_u \rangle
    = \delta_{k0}~,\quad k \in \ZZ_+~.\]
Therefore by Lemma~\ref{l2},
\[ \int_{-d}^d p_k(t) d \langle E_t \delta_u, \delta_u \rangle
    = \int_{-d}^d p_k(t) d\sigma^d_\text{KM}(t) \]
for any $k \in \ZZ_+$, and hence
\[ \int_{-d}^d p(t) d \langle E_t \delta_u, \delta_u \rangle
    = \int_{-d}^d p(t) d\sigma^d_\text{KM}(t) \]
for any polynomial $p$.
\end{proof}

\subsection{Limit theorems for finite graphs}\label{s:fin}

Let $G_n = (V_n, E_n)$ be a sequence of $d$-regular
graphs,
\begin{equation}\label{toinfty}
N_n = \# V_n \underset{n \to \infty}{\longrightarrow} \infty~,
\end{equation}
and let $M_n$ be a symmetric sign matrix on $G_n$.
The following questions arise:
\begin{enumerate}
\item[(a)] Is it true that
\begin{equation}\label{conv}
\mu_{M_n} \longrightarrow \sigma^d_\text{KM}~,
\end{equation}
{\em for every} sequence $M_n$?
\item[(b)] Does (\ref{conv}) hold for $M_n = A(G_n)$?
\item[(c)] Does (\ref{conv}) hold (a.s.) for a {\em random} sequence
$M_n$ (that is, the entries of $M_n$ are random and independent up to the
symmetry assumption,
\[ \PP \{ M_{n,uv} = 1 \} = \PP \{ M_{n,uv} = -1 \} = 1/2, \quad (u,v) \in E~?)\]
\item[(d)] Does the {\em average} spectral measure
$\EE \mu_{M_n}$ (with respect to the random choice of $M_n$ as in (c))
converge to $\sigma^d_\text{KM}$?
\end{enumerate}

It is easy to see that $(a) \Longrightarrow (b)$ and
$(a) \Longrightarrow (c) \Longrightarrow (d)$.
In fact, all the 4 are equivalent.

Denote by $c_k(G)$ the number of closed paths $(u_0, u_1, \cdots, u_k=u_0)$
in $G$, such that $(u_{j-1}, u_{j}) \in E$ for $1 \leq j \leq k$, and
$u_j \neq u_{(j+2) {\hskip-3pt} \mod k}$ for $1 \leq j \leq k$.

If the numbers $c_k(G)$ are small, $G$ looks locally like a tree; hence
the spectral properties of matrices on $G$ should resemble those of
matrices on $H_d$ (cf.\ Proposition~\ref{me}). This is indeed the case;
the following proposition generalises the result of McKay \cite{MK1} on
adjacency matrices (see also Serre \cite{S}).

\begin{prop}\label{prop1}
For every one of the questions (a)-(d), the answer is positive
iff $c_k(G_n)/N_n \to 0$ for $k = 1, 2, \cdots$.
\end{prop}

To prove the proposition, we need some notation. Let
\[\mathfrak{W}_{uv}(k)
    = \mathfrak{W}_{uv}(k, G) = \left\{ (u_0=u, u_1, \cdots, u_k=v)
    \, \mid \, (u_j, u_{j+1}) \in E \right\} \]
be the collection of paths from $u$ to $v$ in $G$.
Consider the subcollection
\[ \widetilde{\mathfrak{W}}_{uv}(k)
    = \left\{ (u_0, \cdots, u_k) \in \mathfrak{W}_{uv}(k)
    \, \mid \, u_j \neq u_{j-2}
    \quad \text{for} \quad j \geq 2 \right\} \]
of {\em non-backtracking} paths, and the subsubcollection
\[ \widetilde{\mathfrak{W}}_{uv}^\text{even}(k)
    \subset \widetilde{\mathfrak{W}}_{uv}(k) \]
of paths on which every edge appears an even number of times.

Finally, denote
\[\begin{cases}
\mathfrak{W}(k, G)
    &= \bigcup_{u \in V} \, \mathfrak{W}_{uu}(k, G), \\
\widetilde{\mathfrak{W}}(k, G)
    &= \bigcup_{u \in V} \, \widetilde{\mathfrak{W}}_{uu}(k, G), \\
\widetilde{\mathfrak{W}}^\text{even}(k, G)
    &= \bigcup_{u \in V} \, \widetilde{\mathfrak{W}}^\text{even}_{uu}(k, G)~.
\end{cases}\]

\begin{lemma}\label{mainlemma}
Let $G = (V, E)$ be a $d$-regular graph and let $p_k = p_{k,d}$
be defined as in (\ref{pk}).
\begin{enumerate}
\item For any symmetric sign matrix $M$ on $G$, and any $u, v \in V$,
\begin{equation}\label{mainall1}
p_k(M)_{uv}
    = \langle p_k(M) \delta_u , \, \delta_v \rangle =
    \frac{\textstyle{\sum^\ast} M_{u_0 u_1} M_{u_1 u_2} \cdots M_{u_{k-1} u_k}}
         {\sqrt{d} \, (d-1)^{(k-1)/2}}~,
\end{equation}
where the sum is over $(u_0, u_1, \cdots, u_k) \in \widetilde{\mathfrak{W}}_{uv}(k)$.
\item In particular,
\begin{equation}\label{mainall}
\left| \langle p_k(M) \delta_u , \, \delta_u \rangle \right|
    \leq \frac{\# \, \widetilde{\mathfrak{W}}_{uu}(k)}
              {\sqrt{d} \, (d-1)^{(k-1)/2}}~,
\end{equation}
with equality for $M = \pm A(G)$.
\item For a randomly chosen $M$,
\begin{equation}\label{maineven}
\EE \langle p_k(M) \delta_u , \, \delta_u \rangle
    = \frac{\# \, \widetilde{\mathfrak{W}}_{uu}^\text{even}(k)}
           {\sqrt{d} \, (d-1)^{(k-1)/2}}~.
\end{equation}
\end{enumerate}
\end{lemma}

\begin{proof}\hfill
\begin{enumerate}
\item For $k=1$, the statement is trivial. Next,
\begin{equation}\label{e1}\begin{split}
p_2(M)_{uv}
    &= \frac{1}{\sqrt{d(d-1)}} \left( M^2 - d \1 \right)_{uv} \\
    &= \begin{cases}
        \frac{1}{\sqrt{d(d-1)}} \sum_w M_{uw} M_{wv}, &u \neq v \\
        \frac{1}{\sqrt{d(d-1)}} \left(\sum_w M_{uw}^2 - d\right) = 0, &u = v
    \end{cases}.
\end{split}\end{equation}
On the other hand,
\[ \widetilde{\mathfrak{W}}_{uv}(k)
    = \begin{cases}
        \left\{ (u, w, v) \, \mid \, (u, w), (w,v) \in E \right\},
            & u \neq v \\
        \varnothing,
            & u = v;
       \end{cases} \]
therefore the right-hand side of (\ref{mainall1}) for $k=2$ is equal to
the right-hand side of (\ref{e1}).

Now proceed by induction.

\item Follows immediately from 1.

\item Take the expectation of both sides of (\ref{mainall1}) and observe that
if $\mathfrak{s_1, \cdots, s_k}$ are random signs drawn with replacement from
a collection $\mathfrak{S}$ of independent random signs, then
\[ \EE \mathfrak{s_1 s_2 \cdots s_k}
    = \begin{cases}
        1, &\begin{aligned}&\text{every term $\mathfrak{s \in S}$ appears an even number} \\
                           &\quad \text{of times in the product ($0$ is even!)}\end{aligned} \\
        0, &\text{otherwise.}
    \end{cases} \]
\end{enumerate}
\end{proof}

Recall the following fact (cf.\ Feller \cite[Ch. VIII, \S 6]{Fe}):
\begin{prop*}
Let $(\mu_n)$ be a sequence of probability measures such that
\[ \int x^k d\mu_n(x) \To \int x^k d\mu(x), \quad k=1,2,3,\cdots~,\]
where $\mu$ is a probability measure with compact support.
Then
\[ \mu_n \To \mu~.\]
\end{prop*}

Now Proposition~\ref{prop1} follows from the next lemma:

\begin{lemma}
Let $G_n = (V_n, E_n)$ be a sequence of $d$-regular
graphs,
\[ \# V_n \underset{n \to \infty}{\longrightarrow} \infty~.\]
The following are equivalent:
\begin{enumerate}
\item For any $k \in \NN$,
\[\# \, \widetilde{\mathfrak{W}}(k, G_n) / \# V_n
    \longrightarrow 0 \]
as $n \to \infty$.
\item For any $k \in \NN$,
\[\# \, \widetilde{\mathfrak{W}}^\text{even}(k, G_n) / \# V_n
    \longrightarrow 0~.\]
\item For any $k \in \NN$,
\[ c_k(G_n) / \# V_n \longrightarrow 0~.\]
\end{enumerate}
\end{lemma}

\begin{proof}
First,
$\widetilde{\mathfrak{W}}^\text{even}(k, G)
    \subset \widetilde{\mathfrak{W}}(k, G)$; hence
\[ \# \, \widetilde{\mathfrak{W}}^\text{even}(k, G)
    \leq \# \, \widetilde{\mathfrak{W}}(k, G) \]
and $1 \Longrightarrow 2$. Similarly,
$c_k(G) \leq \widetilde{\mathfrak{W}}^\text{even}(2k, G)$
(just concatenate a closed path to itself), and so
$2 \Longrightarrow 3$. Finally,
\[ \# \, \widetilde{\mathfrak{W}}(k, G)
    = c_k(G) + \sum_{1 \leq r < k/2} (d-2)(d-1)^{r-1} c_{k-2r}(G)~;\]
therefore $3 \Longrightarrow 1$.

\end{proof}

\section{Spectral measure: proofs}

\subsection{McKay's theorem}\label{s:mk}

Let $(G_n)$ be a sequence of random $d$-regular graphs: $G_n$ is chosen
uniformly from the collection of all $d$-regular graphs on $n$ vertices;
let $M_n$ be a symmetric sign matrix on $G_n$.

\begin{prop*} For any $k \in \NN$, $c_k(G_n) \To 0$ in distribution
as $n \to \infty$.
\end{prop*}
This proposition was first proved by Wormald; see also McKay, Wormald
and Wysocka \cite{MWW} and the discussion below.

\begin{cor} Let $\overline{M}_n$ be an $n \times n$ symmetric $\pm1$
matrix, $n = 1,2,\cdots$. If
\[ M_n = \overline{M}_n \bullet A(G_n) \]
is the {\em Hadamard product} of $\overline{M}_n$ and $A(G_n)$,--
that is,
\[ M_{n,uv} = \overline{M}_{n,uv} A(G_n)_{uv}~,-\]
then
\[ \mu_{M_n} \To \sigma_\text{KM}\]
weakly, in distribution, as $n \To \infty$.
\end{cor}

In particular (for $\overline{M}_{n,uv} = 1$, $1 \leq u,v \leq n$),
we recover McKay's theorem formulated in Subsection~\ref{f:mk};
this is very similar to the original proof in \cite{MK1}.

Now we aim for an estimate on the rate of convergence.

\begin{lemma}\label{cms1}
Let $\mu$ be a probability measure on $\RR$ such that
\begin{equation}\label{3}
\left| \int p_{k,d} \, d\mu \right|
    \leq \eps_k, \quad 1 \leq k \leq 2m-2 \, \text{.}
\end{equation}
Then
\[ d_\text{K} (\mu, \sigma_\text{KM}^d)
    \leq C \left( 1/m + m^6 \sqrt{\sum \eps_k^2}\right)~, \]
where $C>0$ is a universal constant.
\end{lemma}

The case $\eps_1 = \cdots = \eps_{2m-2} = 0$ follows from
the Chebyshev--Markov--Stieltjes inequalities (cf.\ Akhiezer \cite{A});
we present the proof of the general case in Subsection~\ref{s:cms}
(see Proposition~\ref{cms} and Remarks~\ref{r:cms},\ref{r:resc}).

\begin{dfn}
The {\em girth} $\gamma(G)$ of a graph $G$ is the size of the smallest
closed cycle in $G$. In other words,
\[ \gamma(G) = \min \{k \, \mid \, c_k(G) > 0\}~.\]
\end{dfn}

The following proposition was proved by McKay \cite{MK1} with a slightly
weaker estimate, and later by Li and Sol\'e \cite{LS} using the argument
that we reproduce here.

\begin{prop}[McKay, Li--Sol\'e]
Let $G$ be a $d$-regular graph. Then
\[ d_\text{K}(\mu_{A(G)}, \sigma^d_\text{KM}) \leq \frac{C'}{\gamma(G)}~,\]
where $C'>0$ is a universal constant.
\end{prop}

\begin{proof}
By Lemma~\ref{mainlemma},
\[\begin{split}
\int p_k d\mu_{A(G)}
    &= \sum p_k(\lambda_i(A(G))) / n \\
    &= \tr p_k(A)/n
    = \# \widetilde{\mathfrak{W}}(k, G)
        \Big/ \left( n \sqrt{d} (d-1)^{(k-1)/2} \right) = 0
\end{split}\]
for $1 \leq k < \gamma(G)$. Therefore by Lemma~\ref{cms1} (with
all $\eps_k$ equal to $0$)
\[ d_\text{K}(\mu_{A(G)}, \sigma^d_\text{KM}) \leq \frac{C}{\gamma(G)/2}~.\]
\end{proof}

\begin{rmk} Obviously, the last proposition is valid for any symmetric
sign matrix $M$ on $G$.
\end{rmk}

Unfortunately, the girth of a (typical) random $d$-regular graph is
$O(1)$; therefore the proposition is not applicable. To obtain a meaningful
bound in McKay's theorem for random graphs, we use the full strength of
Lemma~\ref{cms1}, as well as the estimates on $\# \mathfrak{W}(k, G)$
that can be extracted from the work of McKay, Wormald and Wysocka \cite{MWW}.
We omit the details that lead to

\begin{prop}
Let $G$ be a random $d$-regular graph on $n$ vertices. Then
\[ d_\text{K}(\mu(A(G)), \sigma^d_\text{K})
    \leq C \sqrt{\frac{\log d}{\log n}}\]
with probability $1 - o(1)$ (as $n \to \infty$), where $C>0$
is a constant independent of $d$ and $n$. Moreover, with
probability $1 - o(1)$,
\[ d_\text{K}(\mu(M), \sigma^d_\text{K})
    \leq C \sqrt{\frac{\log d}{\log n}}\]
for all sign matrices $M$ on $G$ (simultaneously).
\end{prop}

\subsection{Wigner's law}\label{s:w}

Let $A$ be a random $n \times n$ matrix, as in (\ref{d:w}). Then
\[ A = \widetilde{A}/\sqrt{n} + D~,\]
where $\widetilde{A}$ is a random symmetric sign matrix on the
complete graph $K_n$ (every two vertices are connected by an edge),
and $D$ is a diagonal matrix,
\begin{equation}\label{dnorm}
\| D \| = 1/(2\sqrt{n})~.
\end{equation}

We will show that
\[ \mu_A \overset{(4)}{\approx} \mu_{\widetilde{A}/\sqrt{n}}
         \overset{(3)}{\approx} \mu_{\widetilde{A}/\sqrt{n-1}}
         \overset{(2)}{\approx} \widetilde{\sigma}^{n-1}_\text{KM}
         \overset{(1)}{\approx} \sigma_\text{W}~,\]
where $\widetilde{\sigma}^d_\text{KM}$ is the Kesten--McKay
measure scaled to $[-1, 1]$:
\[ d\widetilde{\sigma}^d_\text{KM} (x)
    = d\sigma^d_\text{KM} (2 \sqrt{d-1} \, x)
    = \frac{2d(d-1)}{\pi} \frac{\sqrt{1-x^2} \, dx}{d^2 - 4(d-1)x^2}~.\]

{\bf Step 1:} Let $d \geq 3$. Then
\[\begin{split}
d_\text{K}(\widetilde{\sigma}^{d}_\text{KM}, \sigma_\text{W})
    &\leq \int_{-1}^1  \left|
        \frac{2d(d-1)}{\pi} \frac{\sqrt{1-x^2}}{d^2 - 4(d-1)x^2}
            - \frac{2}{\pi} \sqrt{1-x^2}
        \right| dx \\
    &= \int_{-1}^1 \left| \frac{d(d-1)}{d^2 - 4(d-1)x^2} - 1 \right|
        \times \frac{2}{\pi} \sqrt{1-x^2} \, dx \\
    &\leq \int_{-1}^1 \frac{|d - 4(d-1)x^2|}{d^2 - 4(d-1)x^2} \,
        \times \frac{2}{\pi} \sqrt{1-x^2} \, dx \\
    &\leq \frac{3d}{(d-2)^2} \leq C/d
\end{split}\]
for some universal constant $C>0$.

In particular,
\[ d_\text{K}(\widetilde{\sigma}^{n-1}_\text{KM}, \sigma_\text{W})
    \leq C_1/n~.\]

{\bf Step 2:} Observe that
\[ d_\text{K}(\mu_{\widetilde{A}/\sqrt{n-1}},
        \widetilde{\sigma}^{n-1}_\text{KM})
    = d_\text{K}(\mu_{\widetilde{A}}, \sigma^{n-1}_\text{KM})~.\]
Now we are in the familiar setting of symmetric sign matrices on a graph.

First consider the average spectral measure $\EE \mu_{\widetilde{A}}$.
By Lemma~\ref{cms1},
\begin{multline}\label{e:st2}
d_\text{K}(\EE \mu_{\widetilde{A}}, \sigma^{n-1}_\text{KM}) \\
    \leq C \left( 1/m + m^6 \sqrt{\sum_{k=1}^{2m-2}
        \left[ \int p_{k, n-1}(x) d\EE\mu_{\widetilde{A}}(x) \right]^2}
    \right)~;
\end{multline}
we will take $m = c n^{1/10}$. By Lemma~\ref{mainlemma},
\[\begin{split}
\int p_{k, n-1}(x) d\EE\mu_{\widetilde{A}}(x)
    &= \sum_{u=1}^n \EE \langle p_{k,n-1}(A) \delta_u , \, \delta_u \rangle / n \\
    &= \# \, \widetilde{\mathfrak{W}}^\text{even}(k, K_n)
        \Big/ \left(n \sqrt{n-1} \, (n-2)^{(k-1)/2}\right)~.
\end{split}\]

Obviously, $\widetilde{\mathfrak{W}}^\text{even}(k, K_n) = \varnothing$  for odd $k$,
whereas for even $k$
\[ \# \, \widetilde{\mathfrak{W}}^\text{even}(k, K_n)
    \leq C k  n^{k/2}/4 \leq C k n^{k/2}\]
by Proposition~\ref{comb_kn} (that we prove in Subsection~\ref{s:cntc}). Hence
\begin{equation}\label{estint1}
0 \leq \int p_{k, n-1}(x) d\EE\mu_{\widetilde{A}}(x)
    \leq C k/n~.
\end{equation}

By (\ref{e:st2}), we have proved that
\begin{equation}
d_\text{K}(\EE \mu_{\widetilde{A}}, \sigma^{n-1}_\text{KM})
    \leq C_1 /n^{1/10}
\end{equation}
and therefore
\begin{multline}\label{est2}
d_\text{K}(\EE \mu_{\widetilde{A}/\sqrt{n-1}}, \sigma_\text{W})
    \leq d_\text{K}(\EE \mu_{\widetilde{A}/\sqrt{n-1}}, \widetilde{\sigma}^{n-1}_\text{KM})
        + d_\text{K}(\widetilde{\sigma}^{n-1}_\text{KM}, \sigma_\text{W}) \\
    = d_\text{K}(\EE \mu_{\widetilde{A}}, \sigma^{n-1}_\text{KM})
        + d_\text{K}(\widetilde{\sigma}^{n-1}_\text{KM}, \sigma_\text{W})
    \leq C_2 /n^{1/10}~.
\end{multline}

{\bf Steps 3 and 4:} It remains to recall (\ref{dnorm}) and deduce
\begin{prop} There exists a universal constant $C$ such that,
for a random matrix $A$ defined by (\ref{d:w}),
\begin{equation}\label{est1}
d_\text{K}(\EE \mu_{A}, \sigma_\text{W}) \leq C /n^{1/10}~.
\end{equation}
\end{prop}

With some more effort, it is possible to prove a slightly stronger
proposition:
\begin{prop} There exists a universal constant $C$ such that,
for a random matrix $A$ defined by (\ref{d:w}),
\begin{equation}\label{est3}
d_\text{K}(\mu_{A}, \sigma_\text{W}) \leq C /n^{1/10}~.
\end{equation}
with probability $1 - o(1)$ (as $n \to \infty$).
\end{prop}

\begin{rmk}
G\"otze and Tikhomirov proved \cite{GT} that the left-hand sides of
both (\ref{est1}) and (\ref{est3}) are not greater than $C/\sqrt{n}$;
however, their argument is much more intricate.
\end{rmk}

\subsection{Marchenko--Pastur law}\label{s:mp}

Let $B$ be a random $n \times N$ matrix, as in (\ref{d:mp}). Define an
$(n+N) \times (n+N)$ matrix $\widetilde{B}$ in the following way:
\[ \widetilde{B} =
\left(\begin{matrix}
    0 &\sqrt{N}B^t \\
    \sqrt{N}B &0
\end{matrix}\right)~. \]
Then $\widetilde{B}$ is a symmetric sign matrix on the complete bipartite
graph $K_{n,N}$:
\[\begin{split}
V_{n,N} &= \{1',\cdots,n',1'',\cdots,N''\}, \\
E_{n,N} &= \{ (u', v'') \, \mid 1 \leq u \leq n, 1 \leq v \leq N \}~.
\end{split}\]

The graph $K_{n,N}$ is not regular (unless $n = N$); however, it is bi-regular
(of bi-degree $(N,n)$).
\begin{dfn}
A graph $G = (V' \cup V'', E)$ is called bi-regular (of bi-degree $(d',d'')$) if
\begin{enumerate}
\item $E \subset V' \times V''$
\item The degree of every vertex $v' \in V'$ equals $d'$, and the degree
of every vertex $v'' \in V''$ equals $d''$.
\end{enumerate}
\end{dfn}

Li and Sol\'e proved \cite{LS} an analogue of Lemma~\ref{mainlemma} for bi-regular
graphs and used it to recover the spectral measure of the bi-regular tree
(first computed by Godsil and Mohar \cite{GM}), and to show that the spectral
measure is not far from it for finite bi-regular graphs of large girth, and
for random bi-regular graphs. Here we focus on the limiting case $n, N \To \infty$.

Let
\[ \xi_1 = (n-2)/N, \quad\xi_2 = (n-1)(N-1)/N^2~;\]
note that $\xi_1, \xi_2 \To \xi$ under the assumptions of the Marchenko-Pastur
theorem. Define a sequence of polynomials $q_k = q_{k,\xi_1,\xi_2}$:
\[\begin{split}
&q_0(t) = 1, \quad q_1(t) = (t - 1)/\sqrt{\xi_2}, \\
&q_{k+1}(t) = (t - 1 - \xi_1) q_k(t) / \sqrt{\xi_2} - q_{k-1}(t)~.
\end{split}\]

\begin{lemma}\label{mainbr}\hfill
\begin{enumerate}
\item\label{b1} The polynomials $q_k$ are orthogonal with respect to a certain
(explicit) measure $\sigma^{n,N}_{GM}$ supported on
\[ [1 - 2 \sqrt{\xi_2} + \xi_1, 1 + 2 \sqrt{\xi_2} + \xi_1]~.\]
\item\label{b2} If $n,N \To \infty$ so that $n/N \to \xi$, the measure converges weakly
to the Marchenko--Pastur measure $\sigma^\xi_\text{MP}$. Moreover,
\[ d_\text{K} \left( \sigma^{n,N}_{GM}, \sigma^\xi_\text{MP}\right)
        \leq C/n~.\]
\end{enumerate}
\end{lemma}

\begin{proof}[Sketch of proof]
Both facts can be deduced from an explicit formula for $\sigma^{n,N}_{GM}$,
that follows from Bernstein--Szeg\H{o} formul{\ae} in Subsection~\ref{s:bs}
(cf.\ Li and Sol\'e \cite{LS}).
\end{proof}

\begin{rmk}
For fixed $k$, \[ q_{k,\xi_1,\xi_2} \To q_{k, \xi,\xi}~;\]
$q_{k,\xi,\xi}$ are orthogonal with respect to $\sigma_\text{MP}^\xi$
according to Example~\ref{exMP} in Subsection~\ref{s:bs}. Therefore
the convergence in \ref{b2} can be seen without writing the explicit
formul{\ae} for $\sigma^{n,N}_{GM}$.
\end{rmk}

The following lemma is an analogue of Lemma~\ref{mainlemma}; the proof is
analogous.

\begin{lemma}\label{mainbr1}
If $M$ is an $n \times N$ matrix the entries of which are equal
to $\pm 1$, then
\begin{equation*}
q_k(M M^t/N)_{uv}
    = \frac{\textstyle{\sum^\ast} M_{u_0' u_1''} M_{u_1' u_1''}
        M_{u_1' u_2''} M_{u_2' u_2''} \cdots M_{u_{k-1}' u_k''} M_{u_k' u_k''}}
           {(nN)^{k/2}}~,
\end{equation*}
where the sum is over
\[ (u_0', u_1'', u_1', u_2'', \cdots, u_{k-1}', u_k'', u_k')
    \in \widetilde{\mathfrak{W}}_{uv}(2k, K_{n,N})~.\]
\end{lemma}

Now,
\[\begin{split}
\int q_k d\EE\mu_C
    &= n^{-1} \tr q_k(C) \\
    &= \sum_{u=1}^n \# \widetilde{\mathfrak{W}}_{u'u'}(2k, K_{n,N})
        \Big/ \left(n (nN)^{k/2} \right) \\
    &< \# \widetilde{\mathfrak{W}}(2k, K_{n,N})
        \Big/ \left(n (nN)^{k/2}\right)~.
\end{split}\]

For $k \leq c \xi^{3/20} n^{1/10}$, the last quantity is bounded by
\[ C k /n  \]
according to Proposition~\ref{comb_knN}.

Proceeding as in the previous subsection, with the general
Proposition~\ref{cms} (and the following remarks)
instead of Lemma~\ref{cms1}, we can deduce the following form
of the Marchenko--Pastur theorem:

\begin{prop} Under the assumptions of the Marchenko--Pastur theorem,
\[ d_\text{K} (\EE \mu_C, \sigma^\xi_\text{MP})
    \leq C/\left( \xi^{3/20} n^{1/10}\right)~;\]
moreover,
\[ d_\text{K} (\mu_C, \sigma^\xi_\text{MP})
    \leq C'/\left(\xi^{3/20} n^{1/10}\right)\]
with probability $1 - o(1)$.
\end{prop}

\begin{rmk}
For $\xi$ bounded away from $1$, G\"otze and Tikhomirov proved \cite{GT2}
a better estimate $C / n^{1/2}$ for the left-hand sides in these
inequalities.
\end{rmk}

\section{Extremal eigenvalues}\label{S:norm}

\subsection{Preliminaries}

In the previous sections, the convergence of the spectral measure
$\mu_{A_n} \To \sigma$ followed from the convergence
\begin{equation}\label{e:conv}
\int P_k d\mu_{A_n} \To 0, \quad k = 1,2,3...~,
\end{equation}
where $P_k$ are the orthogonal polynomials with respect to $\sigma$.

To obtain convergence, we only needed (\ref{e:conv}) to hold
for (every) fixed $k$. However, in some of the examples,
the integral on the left-hand side of (\ref{e:conv}) is small also
for $k$ growing with $n$. If this is the case (for $k$ growing
fast enough), no eigenvalues of $A$ can lie far from the support of
$\sigma$. We formalise this observation in this section.

Bai and Yin \cite{BY} applied a similar method (in implicit form)
for random covariance matrices. In \cite{AFMS}, exponentially
decaying  estimates on the probability of deviations were obtained
for this case, using the method Bai and Yin and a formalism similar
to that of the present note. In particular, Subsection~\ref{s:by}
reproduces some of the results in \cite{AFMS} (correcting minor
errors and misprints).

\subsection{The F\"uredi-Koml\'os theorem}

Let $A$ be a random matrix defined as in (\ref{d:w}). As in the first
paragraph of Subsection~\ref{s:w},
\[ A = \widetilde{A}/\sqrt{n} + D~,\]
where $\widetilde{A}$ is a random sign matrix on the complete graph
$K_n$ and $\|D\| \leq 1 \Big/ 2\sqrt{n}$. Recall the estimate
(\ref{estint1}):
\begin{equation*}
0 \leq \EE \sum_{i=1}^n p_{k,n-1}(\lambda_i(\widetilde{A}))
    \leq C k, \quad k \leq c n^{1/10}.
\end{equation*}
By Chebyshev's inequality,
\begin{equation}\label{cheb1}
\PP \left\{ \sum_{i=1}^n p_{k,n-1}(\lambda_i(\widetilde{A}))
    \geq L \right\} \leq Ck/L, \quad L \geq 0~.
\end{equation}

Now, $p_{k, n-1}$ are orthogonal with respect to the measure
$\sigma^{n-1}_\text{KM}$ supported on $[-2 \sqrt{n-1}, 2\sqrt{n-1}]$.
Therefore, for large $k$, $p_k$ tend to infinity very fast outside
this interval. More formally, we have the following

\begin{lemma}\label{pkbound}
There exists a universal constant $C > 0$ such that the inequalities
\begin{enumerate}
\item $\inf_{t \in \RR} p_k(t) \geq - C k$;
\item $\inf_{|t| \geq 2 \sqrt{n-1}(1+\eps)} p_k(t)
    \geq \exp( C^{-1} k \sqrt{\eps})$
\end{enumerate}
hold for any even $k \geq 2$ and any $0 \leq \eps \leq 1$.
\end{lemma}

These estimates follow from the formul\ae{} in Example~\ref{exKM},
combined with (\ref{defCheb}).

Suppose $\widetilde{A}$ has at least one eigenvalue outside
\[ (-2\sqrt{n-1}(1+\eps), 2\sqrt{n-1}(1+\eps))~, 
    \quad \eps \geq C_2 \log^2 n / k^2~. \]
Then, by the above lemma,
\[ \sum_{i=1}^n p_{k,n-1}(\lambda_i(\widetilde{A}))
    \geq \exp( C^{-1} k \sqrt{\eps}) - C (n-1) k
    \geq \exp( C_1^{-1} k \sqrt{\eps})~. \]
According to (\ref{cheb1}), the probability of this event is at most
\[C k \exp(-C_1^{-1} k \sqrt{\eps})
    \leq \exp(-C_3^{-1} k \sqrt{\eps})~.\]
Taking $k = 2 \lfloor c n^{1/10}/2 \rfloor$ and recalling (\ref{dnorm}),
we obtain the following quantitative form of the F\"uredi--Koml\'os theorem:
\begin{thm}\label{fk+}
Let $A$ be a random symmetric $n \times n$ matrix (as in (\ref{d:w})); let
\[ C \log^2 n / n^{1/5} \leq \eps \leq 1~.\]
Then
\begin{equation}\label{e:fk+}
\PP \{ \| A \| \geq 1 + \eps \}
    \leq \exp(-C^{-1} n^{1/10} \sqrt{\eps})~;
\end{equation}
here $C > 0$ is a universal constant.
\end{thm}

In particular, we recover Theorem~\ref{t:bs}
with $\alpha_1 = 1/10$, $\alpha_2 = 1/2$, $\alpha_3 = 0.0999$.

General concentration results yield an improvement $\alpha_1 = 1$,
$\alpha_2 = 2$; this was brought to our attention by Michel Ledoux.
The formal argument is as follows:
\begin{thm}\label{fk++}
Let $A$ be a random symmetric $n \times n$ matrix (as in (\ref{d:w})); let
\[ C_1 \log^2 n / n^{1/5} \leq \eps \leq 1~.\]
Then
\begin{equation}\label{e:fk++}
\PP \{ \| A \| \geq 1 + \eps \}
    \leq \exp(-C_1^{-1} n \eps^2)~;
\end{equation}
here $C_1 > 0$ is a universal constant.
\end{thm}

\begin{proof}
By (\ref{e:fk+}) with $\eps = C \log^2 n/n^{1/5}$, the median
of $\| A \|$ is rather close to $1$:
\[ \operatorname{Med} \| A \| \leq 1 + C \log^2 n / n^{1/5}~.\]
Therefore by the result of Alon, Krivelevich and Vu \cite{AKV},
\[ \PP \left\{ \| A \| \geq 1 + C \log^2 n / n^{1/5} + \eps \right\}
    \leq 8 \exp(-n\eps^2/32)~.\]
\end{proof}

\begin{rmk}\label{rmk:alphas1}
The original proof of Boutet de Monvel and Shcherbina \cite{BS}
yields $\alpha_1 = 1/2$, $\alpha_2 = 3/2$, $\alpha_3 = 0.333$.
The estimate (\ref{e:fk++}) with slightly better constants
can be also deduced from a corresponding estimate for Gaussian
matrices.
\end{rmk}

\subsection{Bai--Yin theorem}\label{s:by}

Proceed similarly to the proof of the F\"uredi--Koml\'os theorem.
According to Subsection~\ref{s:mp}
\[ \EE \sum_{i=1}^n q_k (\lambda_i(C))d\mu_C \leq Ck\]
for $k \leq c \xi^{3/20} n^{1/10}$; hence
\[ \PP \left\{ \sum_{i=1}^n q_k (\lambda_i(C))d\mu_C \geq L \right\}
    \leq Ck/L~.\]

Lemma~\ref{pkbound} extends verbatim:
\begin{lemma}\label{qkbound}
There exists a universal constant $C > 0$ such that the inequalities
\begin{enumerate}
\item $\inf_{t \in \RR} q_k(t) \geq - C k$;
\item $\inf_{|t - 1 - \xi_1| \geq 2 \sqrt{\xi_2}(1+\eps)} q_k(t)
    \geq \exp( C^{-1} k \sqrt{\eps})$
\end{enumerate}
hold for any even $k \geq 2$ and any $0 \leq \eps \leq 1$.
\end{lemma}

Now assume $C$ has at least one eigenvalue outside
\[ [(1-\sqrt{\xi})^2 - \eps, (1 + \sqrt{\xi})^2 + \eps]~.\]
Then
\[ \sum_{i=1}^n q_k (\lambda_i(C))d\mu_C
    \geq \exp( C^{-1} k \sqrt{\eps/\xi}) - C_1 kn
    \geq \exp( C_2^{-1} k \sqrt{\eps/\xi}) \]
if $\eps \geq \frac{C_3 \xi \log^2 n}{k}$. The probability of this event
is at most
\[ C_4 k \exp( - C_2^{-1} k \sqrt{\eps/\xi})
    \leq \exp( - C_5^{-1} k \sqrt{\eps/\xi})~.\]

We have thus proved
\begin{thm}
The probability that $C$ has eigenvalues outside
\[ [(1-\sqrt{\xi})^2 - \eps, (1 + \sqrt{\xi})^2 + \eps]\]
is at most
\[ \exp( - C^{-1} \xi^{-7/20} n^{1/10} \eps^{1/2})\]
for
\[ \frac{C \xi^{7/20} \log^2 n}{n^{1/10}} \leq \eps \leq 1~.\]
\end{thm}

In particular, we recover Theorem~\ref{afms} with
$\beta_1 = 1/10$, $\beta_2 = 1/2$, $\beta_3 = 0.0999$.

\begin{rmk}\label{rmk:alphas2} Similarly to the proof of Theorem~\ref{fk++},
general concentration results yield an improvement $\beta_1 = 1$,
$\beta_2 = 2$ in (\ref{eq:afms.2}); this follows from the result
of Meckes \cite{Me}. We are not familiar with a corresponding argument
for (\ref{eq:afms.1}).
\end{rmk}

\section{Bernstein--Szeg\H{o} measures}\label{S:orth}

\subsection{Some formul{\ae}}\label{s:bs}

In this subsection we explain how to compute the orthogonal polynomials
with respect to the measures we encounter. The formul{\ae} we need
follow from some more general formul{\ae}, first proved by
S.~N.~Bernstein and G.~Szeg\H{o} (see Szeg\H{o} \cite[Theorem~2.6]{Sz}).

Recall that the Chebyshev polynomials $U_k(x)$ (of the second kind) are
defined as
\begin{equation}\label{defCheb}
U_k(\cos \theta) = \frac{\sin ((k+1)\theta)}{\sin \theta}~,
    \quad k \in \ZZ~.
\end{equation}
The following recurrent relation is well-known and easy to verify:
\[ 2x U_k(x) = U_{k+1}(x) + U_{k-1}(x)~.\]

\begin{prop*}
Let $\sigma$ be a measure supported on the segment $[-1, 1]$, such that
\[ d\sigma(x)
    = \frac{2}{\pi \gamma^2} \,
        \frac{\sqrt{1 - x^2} \, dx}
             {(\alpha^2 + (1-\beta)^2) + 2\alpha(1+\beta) x + 4 \beta x^2}~,\]
where $\gamma>0$ and $\alpha,\beta \in \RR$ are such that the denominator is strictly
positive on $[-1,1]$. Then the polynomials $P_k(x)$,
\begin{equation}\label{bsz_pol} P_k(x) =
\begin{cases}
    \gamma \big( U_k(x) + \alpha U_{k-1}(x) + \beta U_{k-2}(x) \big), &k > 0 \\
    \frac{\gamma}{\sqrt{1-\beta}} \, \big(U_k(x) + \alpha U_{k-1}(x) + \beta U_{k-2}(x) \big),
        &k = 0~,
\end{cases}
\end{equation}
are orthogonal with respect to $\sigma$:
\[ \int_{-1}^1 P_k(x) P_\ell(x) d\sigma(x) = \delta_{k\ell}, \quad k,l \geq 0~.\]
\end{prop*}

\begin{rmk}
$P_k$ are linear combinations of $U_k$ and hence satisfy
\begin{equation}\label{rec}
2x P_k(x) = P_{k+1}(x) + P_{k-1}(x), \quad k = 2,3,\cdots~.
\end{equation}
\end{rmk}

\begin{ex}\label{exW} If $\alpha = \beta = 0$ and $\gamma = 1$, then
\[ d\sigma(x) = d\sigma_\text{W}(x)
    = \frac{2}{\pi} \sqrt{1-x^2} \, dx \]
is the Wigner measure;
\[ P_k(x) = U_k(x), \quad k = 0, 1, 2, \cdots \]
\end{ex}

\begin{ex}\label{exKM} Let $\alpha = 0$, $\beta = -(d-1)^{-1}$, and
$\gamma = \sqrt{(d-1)/d}$.
Then
\[ d\sigma(x) = d\widetilde{\sigma}_\text{KM}^d(x)
    = \frac{2d (d-1)}{\pi} \,
        \frac{\sqrt{1-x^2}}{d^2 - 4 (d-1) x^2} \, dx \]
is the scaled Kesten--McKay measure;
\[ P_k(x) =
\begin{cases}
1, & k = 0\\
\sqrt{\frac{d-1}{d}} \, U_k(x)
  - \frac{1}{\sqrt{d(d-1)}} \, U_{k-2}(x), &k = 1, 2, 3, \cdots
\end{cases}\]

\begin{rmk}\label{rmkKM}
Note that $p_{k,d}(x) = P_k(x/2\sqrt{d-1})$ (in view of (\ref{rec}),
this is easy to prove by induction). Therefore $p_{k,d}$ are orthogonal
with respect to $\sigma^d_\text{KM}$.
\end{rmk}

\end{ex}

\begin{ex}\label{exMP}
If $\gamma = 1$, $\alpha = \sqrt{y}$, and $\beta = 0$, then
\[ d\sigma(x) = d\widetilde{\sigma}_\text{MP}^{\xi}(x)
    = \frac{2}{\pi} \frac{\sqrt{1-x^2}}{(1+ \xi) + 2\sqrt{\xi} x} \, \, dx \]
is the scaled Marchenko--Pastur probability measure;
\[ P_k(x) = \begin{cases}
1, &k = 0 \\
U_k(x) + \sqrt{\xi} U_{k-1}(x), &k = 1, 2, \cdots
\end{cases}\]
Hence $q_{k,\xi,\xi}$ are orthogonal with respect to $\sigma_\text{MP}^\xi$.
\end{ex}

\subsection{A proposition in the spirit of P.~L.~Chebyshev, A.~A.~Mar\-kov and
T.~J.~Stieltjes}\label{s:cms}

Let $\sigma$ be a probability measure on $[-1, 1]$; let $P_0,P_1,\cdots$ be
the sequence of orthogonal polynomials with respect to $\sigma$, so that
\[ P_k(x) = \gamma_k x^k + \cdots, \quad \gamma_k > 0 \, \text{.} \]

Denote
\[  B_k = \max_{-1 \leq x \leq 1} |P_k(x)|, \quad
    \rho_k(x) = 1 \Big/ \sum_{i=0}^k P_i(x)^2, \quad
    b_k = \max_{-1 \leq x \leq 1} \rho_k(x) \, \text{.} \]

This section is devoted to the proof of the following proposition.

\begin{prop}\label{cms}
Let $\mu$ be a probability measure on $\RR$ such that
\begin{equation}\label{4}
\left| \int P_k d\mu \right|
    \leq \eps_k, \quad 1 \leq k \leq 2m-2 \, \text{.}
\end{equation}
Then
\[ d_\text{K} (\mu, \sigma)
    \leq 2 b_{m-1} + (1 + m^4 b_{m-1}^2 B_m^4) \sqrt{\sum_{k=1}^{2m-2} \eps_k^2}
    \, \text{.} \]
\end{prop}

This proposition is a ``stability version'' of the Chebyshev--Markov--Stieltjes
inequalities (that correspond to $\eps_1 = \eps_2 = \cdots = \eps_{2m-2} = 0$).
We learned some of the ideas in the proof from the work of Nevai \cite{N}.

Several well-known statements are stated further without
proof; these statements are marked with an asterisk. The reader may
find the proofs in the books of Akhiezer \cite[Ch. III]{A} or
Szeg\H{o} \cite[Ch. II]{Sz}.

\begin{rmk}\label{r:cms} For every measure $\sigma$ that we encounter in this note
(or, more formally, for probability measures in the class considered in the previous
subsection),
\[ b_m \leq C/m \quad \text{and} \quad B_m \leq Cm \, \text{.} \]
Therefore for these measures (\ref{4}) implies
\[ d_\text{K}(\sigma,\mu) \leq C \left( 1/m + m^6 \sqrt{\sum \eps_i^2} \right)
    \, \text{.} \]
\end{rmk}

\begin{rmk}\label{r:resc}  Taking $\sigma = \widetilde{\sigma}^d_\text{MK}$ and
scaling, we recover Lemma~\ref{cms1}.
\end{rmk}

\begin{proof}[Proof of Proposition~\ref{cms}]
Let
\[ -1 < \kappa_{1,m} < \kappa_{2,m} < \cdots < \kappa_{m,m} < 1 \]
be the zeros of $P_m$. Choose $1 \leq s \leq m$ and construct two polynomials,
$R$ and $S$, both of degree at most $2m-2$ and such that
\begin{equation}\label{R}\begin{cases}
R(\kappa_{1,m}) = \cdots = R(\kappa_{s,m}) = 1, \\
\qquad      R(\kappa_{s+1,m}) = \cdots = R(\kappa_{m,m}) = 0, \\
R'(\kappa_{1,m}) = \cdots = R'(\kappa_{s-1,m}) \\
\qquad      = R'(\kappa_{s+1,m}) = \cdots = R'(\kappa_{m,m}) = 0,
\end{cases}\end{equation}
and
\begin{equation}\label{S}\begin{cases}
S(\kappa_{1,m}) = \cdots = S(\kappa_{s-1,m}) = 1, \\
\qquad      S(\kappa_{s,m}) = \cdots = S(\kappa_{m,m}) = 0, \\
S'(\kappa_{1,m}) = \cdots = S'(\kappa_{s-1,m}) \\
\qquad =    S'(\kappa_{s+1,m}) = \cdots = S'(\kappa_{m,m}) = 0 \, \text{.}
\end{cases}\end{equation}

\begin{lemma*}[Markov--Stieltjes]\label{rs}
The inequalities
\[ R \geq \1_{(-\infty,\kappa_{s,m}]} \geq \1_{(-\infty,\kappa_{s,m})} \geq S \]
hold.
\end{lemma*}

By the lemma, $\mu(-\infty,\kappa_{s,m}] \leq \int R d\mu$. Expanding
$R = \sum_{k=0}^{2m-2} a_k P_k$ (where $a_k = \int R P_k d\sigma$),
\begin{equation}\label{e:h1}\begin{split}
\int R d\mu &= \sum a_k \int P_k d\mu \\
    &\leq a_0 + \sum_{i=k}^{2m-1} |a_k| \eps_k
    \leq a_0 + \sqrt{\sum_{k=1}^{2m-2} a_k^2}
            \sqrt{\sum_{k=1}^{2m-2} \eps_k^2} \, \text{.}
\end{split}\end{equation}
Now,
\begin{equation}\label{e:h2}\begin{split}
\sqrt{\sum_{k=1}^{2m-2} a_k^2}
        &\leq \sqrt{\sum_{k=0}^{2m-2} a_k^2} \\
        &= \sqrt{\int R^2 d\sigma}
        \leq \sqrt{\int \1_{(-\infty,\kappa_{s,m}]}^2 d\sigma}
            + \sqrt{\int (R-S)^2 d\sigma} \, \text{,}
\end{split}\end{equation}
since definitely $R \leq \1_{(-\infty,\kappa_{s,m}]} + (R-S)$.

By (\ref{R}-\ref{S}), $R-S$ is a square of some polynomial $p$ of degree $m-1$;
\[ p(\kappa_{t,m}) = \delta_{st}, \quad1 \leq t \leq m~.\]
Therefore $p = \ell_{s,m}$ is $s$-th Lagrange interpolation polynomial
of order $m$.

\begin{lemma}\label{l:est} For $-1 \leq x \leq 1$,
\[ |\ell_{s,m}(x)| \leq m^2 b_{m-1} B_m^2 \, \text{.} \]
\end{lemma}
\begin{proof}[Proof of Lemma~\ref{l:est}]
We start from an expression for $\ell_{s,m}$ that
the reader may find in Szeg\H{o} \cite[Chapter XIV]{Sz}:
\[ \ell_{s,m}(x)
    = \frac{\gamma_{m-1}}{\gamma_m} \,
        \rho_{m-1}(\kappa_{s,m}) \,
        P_{m-1}(\kappa_{s,m}) \,
        \frac{P_m(x)}{x - \kappa_{s,m}} \, \text{.} \]
Let us estimate the terms one by one. First,
\[ \frac{\gamma_{m-1}}{\gamma_m}
    = \int_{-1}^1 x P_{m-1}(x) P_m(x) d\sigma(x)
    \leq \sqrt{\int P_{m-1}^2 d\sigma} \sqrt{\int P_m^2 d\sigma} = 1 \, \text{.} \]
Then, $\rho_{m-1}(\kappa_{s,m}) \leq b_{m-1}$, $|P_m(\kappa_{s,m})| \leq B_m$. By
the Lagrange mean-value theorem and A.A.Markov's inequality (see for example
Todd \cite{T})
\[ \left| \frac{P_m(x)}{x - \kappa_{s,m}}\right|
    \leq \max_{-1 \leq y \leq 1} |P_m'(y)|
    \leq m^2 \max_{-1 \leq y \leq 1} |P_m(y)| = m^2 B_m \, \text{.} \]
The lemma is proved.
\end{proof}

Now recall the Gauss--Jacobi quadrature formula.

\begin{lemma*}[Gauss--Jacobi quadrature]
For any polynomial $q$ of degree not greater than $2m-1$,
\[ \int p \, d\sigma = \sum_{i=1}^m \rho_{m-1}(\kappa_{i,m}) p(\kappa_{i,m}) \, \text{.} \]
\end{lemma*}

Applying (\ref{e:h1}-\ref{e:h2}), Lemma~\ref{l:est} and the Gauss-Jacobi
quadrature, we obtain:
\[\begin{split}
\mu(-\infty,\kappa_{s,m}]
    &\leq \int R d\sigma + (1 + m^4 b_{m-1}^2 B_m^4) \, \sqrt{\sum \eps_k^2} \\
    &= \sum_{i=1}^s \rho_{m-1}(\kappa_{i,m})
        + (1 + m^4 b_{m-1}^2 B_m^4) \, \sqrt{\sum \eps_k^2} \, \text{.}
\end{split}\]
Similarly,
\[ \mu(-\infty,\kappa_{s,m})
    \geq \sum_{i=1}^{s-1} \rho_{m-1}(\kappa_{i,m})
        - (1 + m^4 b_{m-1}^2 B_m^4) \, \sqrt{\sum \eps_k^2} \, \text{.} \]
The measure $\sigma$ satisfies the assumption (\ref{4}) with $\eps_i = 0$; therefore
\[\sum_{i=1}^{s-1} \rho_{m-1}(\kappa_{i,m}) \leq \sigma(-\infty,\kappa_{s,m})
    \leq \sigma(-\infty,\kappa_{s,m}] \leq \sum_{i=1}^s \rho_{m-1}(\kappa_{i,m})
    \, \text{.} \]
The claim of the proposition follows.
\end{proof}

\section{Counting non-backtracking paths}\label{S:cnt}

This section follows \cite{AFMS} (where walks on the complete bi-partite
graph were considered, cf.\ Subsection~\ref{s:cntbr}); we have corrected
minor errors and misprints.

\subsection{Fragments}

Let $G = (V, E)$ be a graph, and let
\[ \mathfrak{w} = (u^\ast, \cdots)
    \in  \widetilde{\mathfrak{W}}^\text{even}(2k, G)~.\]
Consider
$\mathfrak{w}$ as a set of triples $\{ (u, v, r) | 1 \leq r \leq 2k \}$,
meaning that the $r$th edge of $\mathfrak{w}$ goes from $u \in V$ to $v \in V$.

Divide the edges into 3 classes. If $e \in \mathfrak{w}$ is the first edge
to visit a vertex $v \in V$, we will write $e \in T_1$. More formally,
\[ T_1 = \left\{ (u,v,r) \in \mathfrak{w} \, \mid \,
    \forall r' < r, (u', v', r') \in \mathfrak{w} \Longrightarrow
        v \notin \{ u', v' \} \right\}~.\]

The path $\mathfrak{w}$ is even, therefore for every $e \in \mathfrak{w}$
there will be another edge in $\mathfrak{w}$, coincident with $e$.
Denote
\[ T_2 = \left\{ (u,v,r) \in \mathfrak{w} \, \mid \,
    \exists \, ! \, r' < r, (u,v,r') \in T_1
        \vee (v,u,r') \in T_1 \right\}~.\]
Finally, let $T_3 = \mathfrak{w} \backslash (T_1 \cup T_2)$.

A sequence of vertices $f = (u_1, \cdots, u_\ell)$ ($\ell > 1$) is called
a {\em proto-fragment} of $\mathfrak{w}$ if the following 3 conditions hold:
\begin{enumerate}
\item[(i)] for some $r$
\[ (u_1, u_2, r), (u_2, u_3, r+1), \cdots, (u_{\ell-1}, u_\ell, r+\ell-1) \in T_1~;\]
\item[(ii)] for some $r' (> r)$
\[\begin{cases}
\text{either}
    &(u_1, u_2, r'), (u_2, u_3, r'\!+\!1), \cdots,
        (u_{\ell\!-\!1}, u_\ell, r'+\ell\!-\!1) \in T_2 \\
\text{or}
    &(u_\ell, u_{\ell\!-\!1}, r'), \cdots, (u_3,u_2,r'\!+\!\ell\!-\!2),
        (u_2,u_1,r'\!+\!\ell\!-\!1) \in T_2~;
\end{cases}\]
\item[(iii)] $f$ is maximal with respect to (i)-(ii).
\end{enumerate}
If $f$ is a proto-fragment, $u_1 \neq u^\ast$, we call its suffix
$\bar{f} = (u_2, \cdots, u_\ell)$ a fragment of length $\ell - 1$. If $u_1 = u^\ast$,
we call $f$ itself a fragment of length $\ell$. The vertices on $\mathfrak{w}$ are thereby
divided into $F$ fragments.

\begin{lemma}
$F \leq 2 \# T_3 + 1$.
\end{lemma}

This inequality holds for any graph $G$, as one can easily verify.

\subsection{The complete graph}\label{s:cntc}

\begin{prop}\label{comb_kn}
There exist two constants $C,c > 0$ such that, for $k \leq c n^{1/10}$,
\[ \# \widetilde{\mathfrak{W}}^\text{even}(2k, K_n)
        \leq C k n^k~.\]
\end{prop}

The following lemma is obvious:

\begin{lemma}\label{l:kn}
The number of different fragments of length $\ell$ in $K_n$ is
not greater than $n^\ell$.
\end{lemma}

\begin{proof}[Proof of Proposition~\ref{comb_kn}]
First choose the number $S$ of distinct vertices on $\mathfrak{w}$. Then choose
the lengths of the fragments: this can be done in $\binom{S}{F-1} \leq S^F/F!$
ways. Next, choose the fragments themselves; by Lemma~\ref{l:kn}, this can be done
in $\leq n^S$ ways.

There are $2^F$ possibilities to orient the fragments in $T_2$. Now glue the
oriented fragments onto the path; this can be done in $(2k - 2S + 1)^{2F}$
ways.

Every one of the remaining $2k - 2S$ vertices coincides with one of the $S$
vertices on the fragments. Therefore there are $\leq S^{2k-2S}$ possibilities
to arrange these vertices.

Therefore
\[\begin{split}
\# \widetilde{\mathfrak{W}}^\text{even}(2k, K_n)
    & \leq \sum_{S,F}
        \frac{S^F}{F!} n^S 2^F (2k - 2S + 1)^{2F} S^{2k-2S} \\
    & \leq n^k \sum_{S,F}
        \left\{ \frac{CS (k-S)^2}{F} \right\}^F \left(\frac{S^2}{n}\right)^{k-S}~.
\end{split} \]

Now, $F \leq 2 \# T_3 + 1 = 4k - 4S + 5$; the function $x \mapsto (y/x)^x$ is
increasing on $[0, y/e]$; therefore
\[\begin{split}
\# \widetilde{\mathfrak{W}}^\text{even}(2k, K_n)
    & \leq n^k \sum_{S,F}
        \left( C_1 S (k-S)\right)^{4(k-S)} \left( \frac{S^2}{n}\right)^{k-S} \\
    & \leq n^k \sum_{S,F}
        \left( \frac{C_1 S^6 (k-S)^4}{n} \right)^{k-S} \leq C k n^k
\end{split} \]
for $k \leq cn^{1/10}$.

\end{proof}

\subsection{The complete bipartite graph}\label{s:cntbr}

\begin{prop}\label{comb_knN}
There exists two constants $C,c > 0$ such that, for $k \leq c \xi^{3/20} n^{1/10}$,
\[ \# \widetilde{\mathfrak{W}}^\text{even}(2k, K_{n,N})
        \leq C k (nN)^{k/2}~.\]
\end{prop}

The following obvious lemma replaces Lemma~\ref{l:kn}:
\begin{lemma}\label{l:knN}
The number of different fragments of length $\ell$ in $K_{n,N}$ is
not greater than
\[ 2 \sqrt{N/n} (nN)^{\ell/2}~.\]
\end{lemma}

\begin{proof}[Proof of Proposition~\ref{comb_knN}]
Similarly to the proof of Proposition~\ref{comb_kn},
\[\begin{split}
&\# \widetilde{\mathfrak{W}}^\text{even}(2k, K_{n,N}) \\
&\qquad    \leq \sum_{S,F}
        \frac{S^F}{F!} (2 \sqrt{N/n})^F (nN)^{S/2}  2^F (2k - 2S + 1)^{2F} S^{2k-2S} \\
&\qquad    \leq (nN)^{k/2} \sum_{S,F}
        \left(\frac{CS(k-S)^2}{\sqrt{\xi}F}\right)^F
        \left(\frac{S^2}{\sqrt{nN}}\right)^{k-S} \\
&\qquad    \leq (nN)^{k/2} \sum_{S,F}
        \left(\frac{C_1S(k-S)}{\sqrt{\xi}}\right)^{4(k-S)}
        \left(\frac{S^2}{\sqrt{nN}}\right)^{k-S} \\
&\qquad    \leq (nN)^{k/2} \sum_{S,F}
        \left(\frac{C_1S^6(k-S)^4)}{\xi^{3/2}n}\right)^{4(k-S)} \leq C_2 k (nN)^{k/2}
\end{split} \]
if $k \leq c \xi^{3/20} n^{1/10}$.
\end{proof}

\end{document}